\documentclass[preprint2]{aastex631}

\usepackage{graphicx,color,amsmath,amssymb}
\usepackage{aas_macros}
\usepackage{amsmath}
\usepackage{braket}
\usepackage[utf8]{inputenc}

\newcommand{\nocontentsline}[3]{}
\newcommand{\toclesslab}[3]{\bgroup\let\addcontentsline=\nocontentsline#1{#2\label{#3}}\egroup}
\newcommand{\tocless}[2]{\bgroup\let\addcontentsline=\nocontentsline#1{#2}\egroup}
\newcommand{\epar}{E_\parallel}

\newcommand{\bfE}{{\bf E}}
\newcommand{\bfB}{{\bf B}}

\newcommand{\gagg}{g_{a\gamma\gamma}}

\newcommand{\alfven}{Alfv\'{e}n }
\newcommand{\dB}{\delta \bfB}
\newcommand{\edb}{{\bf E} \cdot {\bf B}}

\newcommand{\kperp}{k_\perp}
\newcommand{\kpar}{k_\parallel}
\newcommand{\jpar}{j^{(1)}_\parallel}
\newcommand{\jpart}{j^{(2)}_\parallel}

\begin{document}

\title{Axion-mediated Transport of Fast Radio Bursts Originating in Inner Magnetospheres of Magnetars}

\author[0000-0001-9115-7844]{Anirudh Prabhu}
\affiliation{Princeton Center for Theoretical Science, Princeton University, Princeton, NJ 08544, USA}

\begin{abstract}


Among magnetar models of Fast Radio Bursts (FRBs), there is ongoing debate about whether the site of coherent radio emission lies within or beyond the light cylinder. We propose a mechanism by which FRBs produced near the magnetar surface are transported out of the magnetosphere by axions, which couple to photons. If the emission site hosts strong accelerating electric fields, a considerable fraction of the FRB energy budget is converted to an axion burst. Once produced, the axion burst free-streams out of the magnetosphere due to the rapidly-decreasing magnetic field. The burst may escape through either the open or closed magnetosphere, while retaining the temporal signature of the original FRB. In the wind region, axions resonantly excite ordinary (O) modes that escape as the plasma density decreases. The radio efficiency of this mechanism satisfies energetics constraints from FRB 121102 for axion-photon coupling strengths that have not been excluded by other astrophysical probes.


\end{abstract}

\section{Introduction} \label{sec:intro}

Fast radio bursts (FRBs) are ultraluminous, millisecond-duration radio transients that are usually of extragalactic origin. Their short durations, with pulse substructure at the microsecond scale, are suggestive of a compact object central engine, such as a neutron star or stellar mass black hole (\citealt{Katz2017, Li2018, Katz2020, Sridhar2021, Katz2022}). Recent detection of an FRB associated with galactic magnetar SGR 1935+2154 provides support for the claim that at least a subset of FRBs are sourced by magnetars. The high brightness temperature (in excess of $10^{30}$ K) necessitates a coherent emission mechanism. Within magnetar models there are two broad categories for the emission: magnetospheric models, in which the radio waves are produced within the light cylinder ($ r \ll R_{\rm LC} \equiv c/\Omega$, where $\Omega$ is the rotational frequency of the magnetar), and far-field models, in which the radio waves are produced outside the light cylinder, possibly in the wind or nebula regions. Examples of magnetospheric models include coherent radiation by charged bunches (\citealt{RudermanSutherland1975, Kumar2017, Kumar2020, Lu2020}) and inner magnetosphere reconnection (\citealt{Lyubarsky2020})\footnote{The author posits that inner-magnetosphere reconnection necessarily takes place, but any FRB produced therein cannot escape.}. Far-field models often rely on magnetosphere ejecta sourced by magnetar flares forming blast waves in the wind that emit coherent radio waves through the synchrotron maser process (\citealt{Beloborodov2017, Metzger2019, Beloborodov2020}) or reconnect with magnetic fields outside the light cylinder, leading to radio emission through the collision of small magnetic plasmoids (\citealt{Philippov2019, Lyubarsky2020, Yuan2020, Mahlmann2022}). Arguments in favor of magnetospheric models are that they are able to explain the rapid variability of FRB pulses, down to 60 ns from FRB 20200120E (\citealt{Nimmo2021}), the rapid polarization swings observed in some FRB pulses (\citealt{Luo2020}), and the relatively high expected radio efficiency of inner magnetosphere mechanisms. One issue that inner magnetosphere models must contend with is how an FRB produced in the inner magnetosphere can escape. As an FRB produced in the inner magnetosphere propagates through the lower density outskirts of the magnetosphere it strongly scatters with the plasma and loses much of its energy to acceleration of charged particles (\citealt{Beloborodov2021}). Models invoking coherent emission by charged bunches circumvent these issues by resorting to emission along open field lines, where radio photons may easily escape (\citealt{Kumar2017, Kumar2020, Lu2020}). Highly beamed emission may be in tension with the high expected event rate of $\sim 10^4$ per day (\citealt{Thornton2013}). Far-field models struggle to explain rapid variability and polarization angle swings, but can successfully explain the escape of radio waves, since the FRB energy is transported out of the magnetosphere by magnetic flare ejecta and do not rely on beamed emission.

In this Letter, we propose a generic class of models in which FRBs produced in the inner magnetosphere are transported out of the magnetosphere by axions. Axions, broadly defined, are ultralight spin-0 bosons that arise in many extensions of the Standard Model (SM) of particle physics. The so-called \emph{QCD axion} was originally proposed to explain the seemingly unnatural smallness of the neutron electric dipole moment (\citealt{PQ1, PQ2, WeinbergAxion, WilczekAxion}). Axion-like particles (ALPs) are also a generic prediction of String Theory (\citealt{Svrcek_2006, Axiverse2010}). Both QCD axions and ALPs are among the best-motivated candidates to account for the $\approx 85$ \% of matter density in the universe that exists in the form of some cold dark matter (CDM) (\citealt{Sikivie1983, Fischler1982}). We use the term ``axion'' to refer to both QCD axions and ALPs, drawing the distinction only when necessary. 

Axions couple to photons through the two-photon interaction term, 

\begin{align}
    \mathcal{L}_a = -\gagg a(x) \edb, \label{eqn:coupling}
\end{align}

\noindent where $a(x)$ is the axion field, $\gagg$ is a coupling constant, and $\bfE$ and $\bfB$ are electric and magnetic fields. This coupling has been exploited in numerous astrophysical and laboratory searches for axions. There is extensive literature on the role of axions in high-energy astrophysical settings, such as in SN1987A (\citealt{Burrows1987, Keil1997, Caputo2022, Hoof2022}). Some constraints from SN1987A come from the fact that axions may efficiently transport energy from the supernova or remnant neutron star, leading to an additional cooling mechanism, analogous to neutrino cooling (\citealt{Burrows1987}). Extreme, highly magnetized plasmas also provide a good laboratory for axion searches. Large magnetic fields, such as those near the surfaces of pulsars and magnetars, greatly enhance the conversion rate between axions and photons. The effects of conversion of axion dark matter in neutron star magnetospheres have been explored in (\citealt{Pshirkov:2007st, Huang:2018lxq, Hook:2018iia, Safdi:2018oeu, Battye:2019aco, Leroy:2019ghm, Foster:2020pgt, Buckley2021, Witte2021, Battye2021, battye2021robust, Millar2021, Foster:2022fxn}). Additionally, there are claims that collisions of axion stars with neutron stars might be the source of FRBs (\citealt{Tkachev2015, Iwazaki2015, Pshirkov2017, Prabhu2020, Buckley2021}). 

 \begin{figure} 
    \includegraphics[width=\columnwidth]{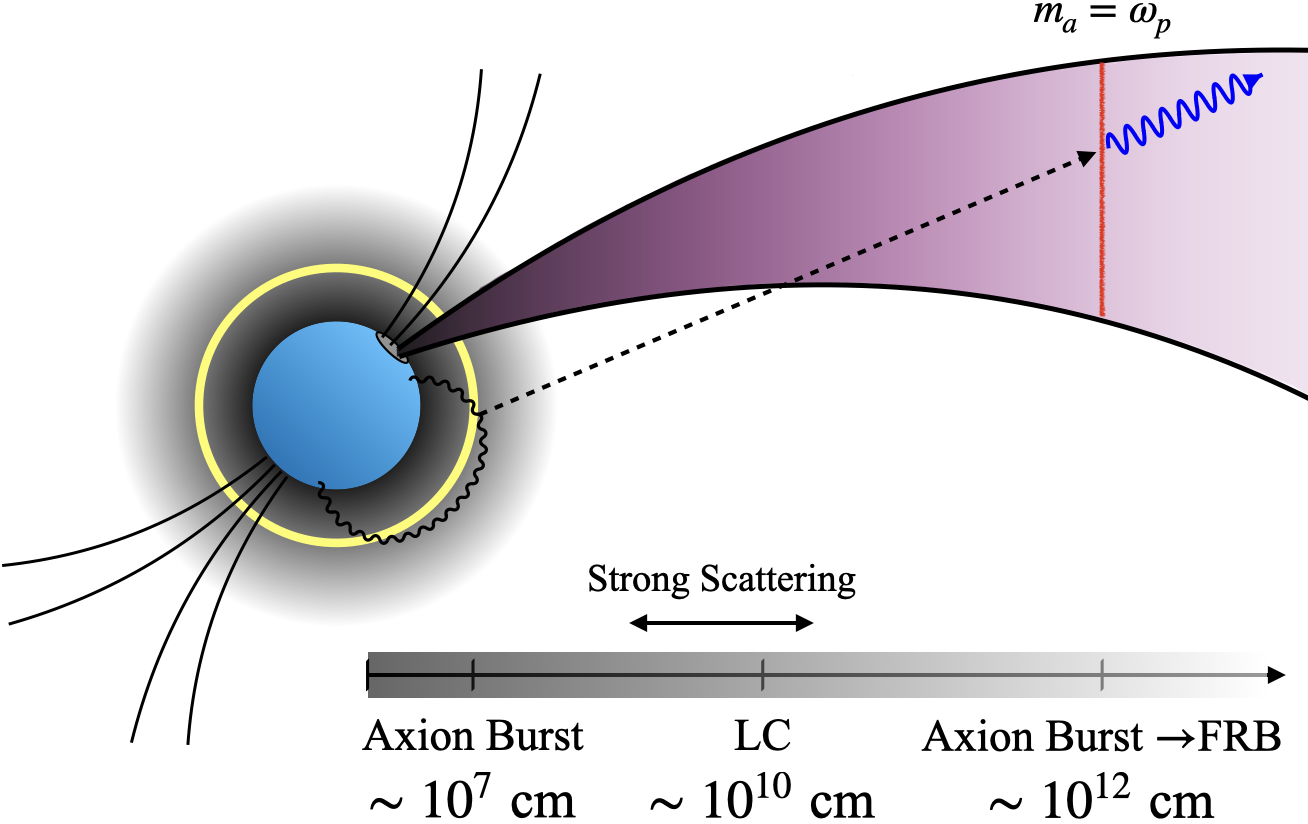}
    \caption{Summary of proposed mechanism. Axions (black, dashed) produced at a critical radius in the inner magnetosphere (yellow circle) propagate through the light cylinder (LC) to the dense, pre-flare wind (purple) and resonantly convert to radio photons (blue) which may escape. Relevant length scales are shown at the bottom of the image for a magnetar with rotational period $P=1$ sec and axion mass $m_a = 10^{-10}$ eV. See Sec. \ref{sec:reconversion} for details of these length scales. The region where an FRB would scatter strongly with the plasma is labeled.  }
    \label{fig:Summary}
\end{figure}

In our model (see Fig.~\ref{fig:Summary} for a summary), an FRB produced near the surface of a magnetar efficiently converts to axions, due to the strong magnetic field therein. Crucially, the axion burst retains the temporal structure of the FRB. As the axion burst propagates through the magnetosphere, its conversion rate back to photons becomes highly suppressed as the magnetic field decreases as $B_\text{dipole} \propto r^{-3}$, allowing the burst to effectively free-stream out of the magnetosphere. Once in the wind region, the axion burst experiences a plasma with decreasing plasma frequency. At a critical radius $r_c \gg R_{\rm LC}$ the plasma frequency coincides with the axion mass, $ \omega_p(r_c) = m_a c^2/\hbar$, and the axion burst resonantly excites ordinary (O) waves which can escape the system. We show that in young magnetars, the radio efficiency of the proposed model is within energetics constraints from FRB 121102, for axion-photon couplings that satisfy high-energy astrophysical constraints. In this model the FRB can escape through either the open or closed magnetosphere and thus does not rely on beaming. 

The paper is organized as follows. In Sec. \ref{sec:axion_photon}, we review the formalism of axion coupling to photons in a magnetized plasma, paying particular attention to regimes of strong coupling. In Sec. \ref{sec:axion_magnetosphere}, we provide some general criteria for axion bursts to be produced in the inner magnetosphere. We relate this discussion to existing magnetospheric FRB models and argue that a large fraction of the FRB energy is converted to axions near the magnetar surface. Following their production, axions re-convert to outgoing radio photons in the enhanced pre-flare magnetar wind. We compute the efficiency of this re-conversion in Sec. \ref{sec:reconversion} and show the axion-photon coupling parameter space consistent with efficiency constraints in Sec. \ref{sec:efficiency}. We provide concluding remarks in Sec. \ref{sec:conclusion}.

\section{Axion-Photon Mixing in Plasma} \label{sec:axion_photon}

Axions interact with electromagnetism through the operator described by equation (\ref{eqn:coupling}). This interaction implies that axions can be generated electromagnetically in regions where $\edb \ne 0$, such as in the vacuum gaps above pulsar polar caps (\citealt{Prabhu2021, Noordhuis2022}). Equation (\ref{eqn:coupling}) also leads to a non-linear modification of Maxwell's equations that allows for interconversion between axions and photons \footnote{For detailed treatments of non-relativistic axion-to-photon conversion in magnetized, anisotropic plasmas, see (\citealt{Millar2021, Witte2021}). We emphasize that these studies consider non-relativistic axion dark matter, thus their results are not directly applicable to the case of relativistic axions presented in this work.} . By assuming 1D propagation in a magnetic field with a constant direction but gradually changing magnitude, and by considering ultrarelativistic axions with energy ($\omega$) much greater than their mass ($m_a$), the equation for axion-photon conversion in a plasma with spatially varying frequency, $\omega_p(z)$, can be derived using the WKB approximation (see, e.g. (\citealt{RaffeltStodolsky1988})),

\begin{align} \label{eqn:schrodinger}
    \left[i \partial_z + \omega \begin{pmatrix}
    1 + \Delta_a & \Delta_{a\gamma} \\
    \Delta_{a\gamma} & 1 + \Delta_\parallel
    \end{pmatrix} \right] \begin{pmatrix}
    a \\
    A_\parallel
    \end{pmatrix} = 0, 
\end{align}

\noindent where $a$ is the axion field, $A_\parallel$ is the photon state parallel to the background magnetic field and normal to the direction of motion ($z$-direction). The diagonal terms represent effective refractive indices with $\Delta_a = -m_a^2/(2\omega^2)$ and $\Delta_\parallel = - \omega_p^2/(2\omega^2)$. \color{black}The refractive index receives a contribution from the Euler-Heisenberg term $\Delta_{\parallel, \text{EH}} = (7\alpha^2/90\pi) (B/B_c)^2 \sin^2 \theta$, where $B_c = m_e^2/e \approx 4.4 \times 10^{13}$ G is the Schwinger critical field (\citealt{RaffeltStodolsky1988}). We will discuss axion production at $\mathcal{O}(10)$ stellar radii, where $\Delta_{\parallel, \text{EH}} \ll \omega_p^2/\omega^2 \sim 1$, so we neglect the Euler-Heisenberg term \color{black}. The off-diagonal term, $\Delta_{a\gamma} = \gagg B_0(z) \sin{\theta}/(2 \omega)$, where $B_0(z)$ is the magnitude of the background magnetic field and $\theta$ is the angle between the magnetic field and the direction of propagation, quantifies the mixing between axion and photon states. Equation (\ref{eqn:schrodinger}) resembles the Schr\"{o}dinger equation, with $z$ playing the role of time, and can be solved to first order in $\Delta_{a\gamma}$ using time-dependent perturbation theory, giving (\citealt{RaffeltStodolsky1988,Fortin2019,Dessert2019}), 

\begin{align} \label{eqn:conversionprobability}
    P = \left|\displaystyle\int_{z_i}^{z_f} dz' 
 \omega \Delta_{a\gamma}(z') e^{i \Delta_a(z') - i \int_{z_0}^{z'} \Delta_\parallel(z'') dz''} \right|^2,
\end{align}

\noindent where $z_i$ and $z_f$ are the initial and final locations, respectively. In general, the dispersion relations for the axion and photon are different, leading to dephasing over a propagation distance $\ell_d = 1/\Delta k$, where $\Delta k = |\sqrt{\omega^2 - \omega_p^2} - \sqrt{\omega^2-m_a^2}|$ is the momentum mismatch between the two modes. If, however, the axion mass coincides with the plasma frequency, the mixing is resonantly enhanced. Axions propagating through a plasma with frequency $\omega_p(z) \propto z^{-n}$, where $n = 3/2$ corresponds to a Goldreich-Julian density profile and $n = 1$ corresponds to a spherically expanding wind will resonantly covert at a critical distance, $z_c$ where $\omega_p(z_c) = m_a c^2/\hbar$. Equation (\ref{eqn:conversionprobability}), can be solved in the stationary phase approximation to give

 \begin{align}\label{eqn:resonantconversion}
    P_\text{res} = \left( \frac{\pi}{n} \right) \gagg^2 B(z_c)^2 \frac{\omega z_c}{\omega_p(z_c)^2 }, 
 \end{align}

\noindent where we have set $\hbar = c = 1$. The analysis above, applied to plane wave solutions, can be extended to generic wave-packets. From (\ref{eqn:resonantconversion}), we note that the conversion leads to spectral distortions, with more efficient conversion of higher frequency components. This is because at fixed axion mass, the momentum mismatch, $\Delta k \approx |\omega_p^2 - m_a^2|/2\omega$, is smaller for higher frequencies. We also note that the axion and photon wave packets experience dispersive spreading. Since in the resonant conversion region, the dispersion relations for axions and photons are roughly the same, this effect is indistinguishable from dispersion of an electromagnetic wave propagating through a plasma.

\section{Axion Production in the Inner Magnetosphere} \label{sec:axion_magnetosphere}

We outline some general criteria for an FRB produced in the inner magnetosphere to convert efficiently into axions. Firstly, a strong electric field component parallel to the background magnetic field ($\epar$) must be present in the emission region. Parallel electric fields arise generically in models of persistent and transient emission in compact objects. Dynamical screening of $\epar$ by pair cascades, leading to the emission of superluminal ordinary (O) modes, is thought to be the mechanism behind pulsar radio emission from polar caps (\citealt{Philippov2020, RudermanSutherland1975}). During a magnetar flare, magnetic reconnection is likely to take place in the inner magnetosphere (\citealt{Lyubarsky2020}) and is  generically expected to host $\epar$ in three dimensions(\citealt{Schindler1988}). Another class of proposed inner magnetosphere models involves the formation of charged bunches that emit coherent curvature radiation (\citealt{Kumar2020,Lu2020}). In this letter, we study axion production within this model, though we emphasize that the general mechanism of axion transport applies to any inner magnetosphere emission mechanism that hosts large $\epar$. We investigate axion production in near-surface magnetic reconnection in future work.

\subsection{Axion Production by Charge-starved \alfven Waves} \label{sec:alfven}

In this section we review the model of (\citealt{Kumar2020}) and compute the efficiency of axion production in this model. The starting point of this model is the release of an enormous amount of magnetic energy during a magnetic flare. This disturbance to the magnetosphere launches \alfven waves from close to the magnetar surface which propagate along the dipole magnetic field lines. The waves require a current along the magnetic field lines, supported by counter-propagating $e^\pm$ pairs accelerated to very high Lorentz factors. The counter-propagating charges are susceptible to a two-stream instability that generates a large $\epar$, and considerable flux of axions. Below, we provide the details of this model.

For simplicity, we describe the propagation of  \alfven waves in a constant magnetic field, $B_0 \hat{z}$, which we will show is a good approximation for the system under consideration. The \alfven wave can be described as a magnetic perturbation in the $\hat{y}$ direction that propagates in the $x-z$ plane, $\dB = B_y \hat{y} \exp[i (k_\perp x + k_\parallel z - \omega_A t )]$, where $B_y = \varepsilon B_0$ and $\varepsilon \ll 1$. For \alfven waves, the perpendicular component of the wave vector decreases with distance from the magnetar surface as $r^{-3/2}$ (\citealt{Lu2020}). Thus $\kperp \ll \kpar$ in general, confirming that a 1D treatment is a good approximation. The amplitude of the \alfven wave also decays with radius due to propagation effects (\citealt{Lu2020}) and the development of the two-stream instability, but the former occurs over large length scales. We work in the approximation in which the two-stream instability and the axion field do not strongly modulate the \alfven wave. This is supported by simulations in (\citealt{Kumar2022}), which show that the maximum amplitude of $\epar$ generated in the two-stream instability is $\sim 0.1 B_y$, and that treating the pre-specified magnetic perturbation defined above to be fixed is a good approximation. At a distance of $\mathcal{O}(10)$ stellar radii, the \alfven wave encounters an upstream medium with low number density. The evolution of $\epar$ and the axion field are described by Amp\`{e}re's law and the Klein-Gordon equation,

\begin{align}
    \partial_t \epar + \jpar - \partial_t \theta B_0 &= \left( \boldsymbol{\nabla} \times \dB \right)_z - \jpart \label{eqn:ampere} \\
    \partial_t^2 \theta - \nabla^2 \theta + m_a^2 \theta &= -\gagg^2 B_0 \epar \label{eqn:KG}
\end{align}

\noindent where $\theta(x) \equiv \gagg a(x)$ is the dimensionless axion field. As in (\citealt{Kumar2022}), we have split the current density into two parts: $\jpar$ is the current density of the upstream medium and $\jpart$ is the current density advected by the \alfven wave. Even in the absence of an upstream medium, a small Coulomb field is generated as the charged particles in the \alfven wave lag behind the wave, leading to a contribution to the RHS of (\ref{eqn:ampere}). The lag between the particles and the \alfven wave grows with time, leading to (\citealt{Kumar2022})

\begin{align} \label{eqn:eparcs}
    \epar(z,t) &= {B_y(z,t) \omega_A t \over 2 \gamma_0^2} \left( {\kperp \over \kpar} \right),
\end{align}

\noindent where $\gamma_0$ is the Lorentz factor of particles advected by the \alfven wave. For axions with mass $m_a \le \omega_A$ the displacement current drives the axion field at frequency $\omega_A$. \alfven waves have a high reflection coefficient at the magnetar surface, allowing them to bounce along field lines in the magnetosphere many times (\citealt{Beloborodov2015}). This may lead to a prolonged, low-frequency axion counterpart at the \alfven wave frequency. We investigate this effect in a future study. 






\subsection{Linear Stability Analysis: Two-stream Instability}

Equations (\ref{eqn:ampere}) and (\ref{eqn:KG}) may be unstable to small perturbations in $\jpart$. A detailed analysis of this two-stream instability was demonstrated analytically and numerically in (\citealt{Kumar2022}). Small parameters, $\epar, \theta$, and $\delta \jpar$ are taken to have dependence $\propto \exp(i k z - i \omega t)$, where $k \in \mathbb{R}$ and $\omega \in \mathbb{C}$. Solutions to the perturbed equations with $\text{Im}(\omega) \equiv \omega_i > 0$ correspond to exponentially growing solutions. The evolution of instabilities will be dominated by the fastest-growing modes which have $\text{Re}(\omega), k \approx \omega_{p,1}$ and $ \omega_i = \omega_{p,1}^{1/3} \omega_{p,2}^{2/3}/\gamma_0$, where $\omega_{p,1(2)} = \sqrt{e^2 n_{1(2)}/m_e}$. The subscript ``1'' corresponds to the upstream plasma and ``2'' to the plasma advected by the \alfven wave (\citealt{Kumar2022}). From (\ref{eqn:KG}), the conversion efficiency from $\epar$ to axions is given by

\begin{align} \label{eqn:pga}
    \eta_{\gamma \to a} &\equiv { \omega^2 |a|^2 \over |\epar|^2} \simeq {\gagg^2 B_0^2 \omega_p^2 \over (m_a^2 + \omega_i^2)^2 + 4 \omega_p^2 \omega_i^2} \\
    & \approx 0.01 \times g_{10}^2 B_{0,13}^2 \omega_{i, -8}^{-2}, \nonumber
\end{align}

\noindent where $g_{10} = \gagg/(10^{-10}/\text{GeV})$, $B_{0,13} = B_0/(10^{13} \text{ G})$, and $\omega_{i, -8}^{-2} = \omega_i/(10^{-8} \text{ eV})$. The fiducial value of $\omega_i$ assumes $\omega_p = 2\pi$ GHz, $\varepsilon = 10^{-3}$, $\kperp = 0.1 \omega_A$, and $\gamma_0 = 1000$. The fiducial value of $\gagg$ is approximately set to the bound from the CAST solar axion search (\citealt{CAST2017}). In (\ref{eqn:pga}), the conversion probability is independent of $m_a$ only if $m_a \ll \omega_p$.

\section{Re-Conversion in the Wind} \label{sec:reconversion}


Magnetars emit a persistent spin-powered wind with luminosity $L_w =  \mu^2 \Omega^4 (1 + \sin^2 \alpha)$, where $\mu$ is the magnetic dipole moment of the magnetar, $\Omega$ its rotational frequency, and $\alpha$ the angle between the rotational and dipole axes (\citealt{Li2012}). The toroidal magnetic field in the wind is supported by a particle outflow, $\dot{N}$. Seconds before a magnetar flare, the wind luminosity and particle flow rates are greatly enhanced from their persistent values (\citealt{Beloborodov2017, Beloborodov2020}). The wind can be characterized by a magnetization $\sigma_w$ and Lorentz factor $\Gamma_w$. \color{black} At large distance the wind Lorentz factor is approximately $\Gamma_w \sim 3 \times (L_w/\dot{N} m_e)^{1/3}$ (\citealt{Beloborodov2020}). In general the wind is highly magnetized, giving $B^2_\phi(r) = L_w/(4\pi r^2)$ at large distance. The plasma frequency can be related to the particle flow rate as $\omega_p^2 = e^2 \dot{N}/(4 \pi m_e r^2 \Gamma_w)$. The critical radius, $r_c$, at which axions resonantly convert to longitudinal O modes, and corresponding magnetic field strength are,

\begin{align}
    r_c &= 2\times 10^{12} \text{ cm} \cdot L_{w,42}^{-1/6} \dot{N}_{40}^{2/3} m_{a,-10}^{-1}, \\
    B_\phi(r_c) &= 100 \text{ G} \cdot L_{w,42}^{-2/3} \dot{N}_{40}^{-2/3} m_{a,-10},
\end{align}

\noindent where $m_{a,-10} = m_a/(10^{-10} \text{ eV})$, $L_{w, 42} = L_w/(10^{42} \text{ erg/sec})$, $\dot{N}_{40} = \dot{N}/(10^{40}/\text{sec})$. For reference, a young magnetar with surface field $B_s = 10^{15}$ G and rotation period $P = 0.1$ sec, would have persistent wind luminosity $L_{w,42} \approx 0.1$ and $\dot{N}_{40} = 10^{-4} \mathcal{M}$, where $\mathcal{M}$ is the multiplicity factor of pairs produced in the open field lines (\citealt{Beloborodov2020}). Again, the pre-flare values of these parameters are expected to be much larger, but are poorly constrained. In Fig. \ref{fig:Exclusion} we adopt a fiducial enhancement factor of $10^2$ for both $L_w$ and $\dot{N}$, but emphasize that the results of this model are sensitive to these parameters and better estimates are required. From (\ref{eqn:conversionprobability}), the resonant axion-to-photon conversion probability in the wind is

\color{black}

\begin{align} \label{eqn:pag}
    \eta_{a \to \gamma} \simeq 0.01 \times g_{10}^2 L_{w,42}^{7/6} \dot{N}_{40}^{-2/3} m_{a,-10}^{-1} \omega_\text{GHz},
\end{align}

\noindent where $\omega_\text{GHz} = \omega/(2\pi \text{ GHz})$. We comment that at arbitrarily low mass $m_a \lesssim 10^{-12}$ eV, resonant conversion takes place outside the wind region and the wind model adopted above does not apply. We do not consider conversion in the nebula region or interstellar medium.

\section{Efficiency Constraints} \label{sec:efficiency}

The energy reservoir for young magnetars comes from either the magnetic field ($E_\text{mag} \simeq {4\pi B_0^2 R_\text{ns}^3/3}$) or rotation ($E_\text{rot} = M_\text{ns} R_\text{ns}^2 \Omega_\text{rot}^2/5$). For the remainder of this section we take standard values for the neutron star mass $(M_\text{ns} = 1.4 \ M_\odot)$ and radius ($R_\text{ns} = 10 \text{ km}$). The energy released in a single burst is 

\begin{align} \label{eqn:efficiency}
    E = {L_\text{iso} \tau \over \eta_r} \left( {\delta \Omega} \over 4\pi \right), 
\end{align}

\noindent where $L_\text{iso}, \tau, \eta_r,$ and the term in the parentheses are the isotropic equivalent luminosity, duration, radio efficiency, and beaming factor of the burst, respectively. Upper limits on the radio energy of several FRBs were derived in (\citealt{Zhang2018Efficiency}), giving $E \eta_r \sim 10^{40}-10^{42}$ erg. The constraint that the total energy released in an FRB not exceed the available energy is,

\begin{align}
    \eta_r &\ge {3 \times 10^{-9} \over (1+z)} f_b
 {D_{L,\text{Gpc}}^2 F_{\nu, \text{Jy} \cdot \text{ms}} \Delta \nu_{\text{GHz}}} B_{0,15}^{-2},
\end{align}

\noindent where $D_{L,\text{Gpc}}$ is the luminosity distance (in Gpc), $F_{\nu, \text{Jy} \cdot \text{ms}}$ is the fluence (in Jy$\cdot$ms), $\Delta \nu_{\text{GHz}}$ is the bandwidth (in GHz), $B_{0,15}$ is the surface magnetic field  (normalized to $10^{15}$ G), $f_b = \delta \Omega /4\pi$, and $z$ is the redshift of the source. The situation is more complicated for repeaters, where the constraint is on the total energy emitted by all bursts. Stringent efficiency constraints can be derived from FRB 121102. A 47-day observation of FRB 121102 with the FAST telescope reported 1652 bursts over 59.5 hours (\citealt{Li2021}), corresponding to an active phase duty cycle of $\zeta \approx 0.05$. The total source energy corresponding to these bursts is $E_\text{tot} \eta_r \simeq {6 \times 10^{42} \text{ erg} \times F_b (\zeta/0.05)^{-1}}$ (\citealt{Zhang2022Review}). Here $F_b = \Delta \Omega/4\pi$ is the global beaming factor, which encompasses the solid angle of all bursts and is in general much greater than $f_b$. The efficiency constraint from FRB 121102 is then

\begin{align}
    \eta_r \gtrsim 1.8 \times 10^{-5} \times F_b B_{0,15}^{-2} \left( {\zeta \over 0.05} \right)^{-1}.
\end{align}

\begin{figure}
    \centering
    \includegraphics[scale=0.40]{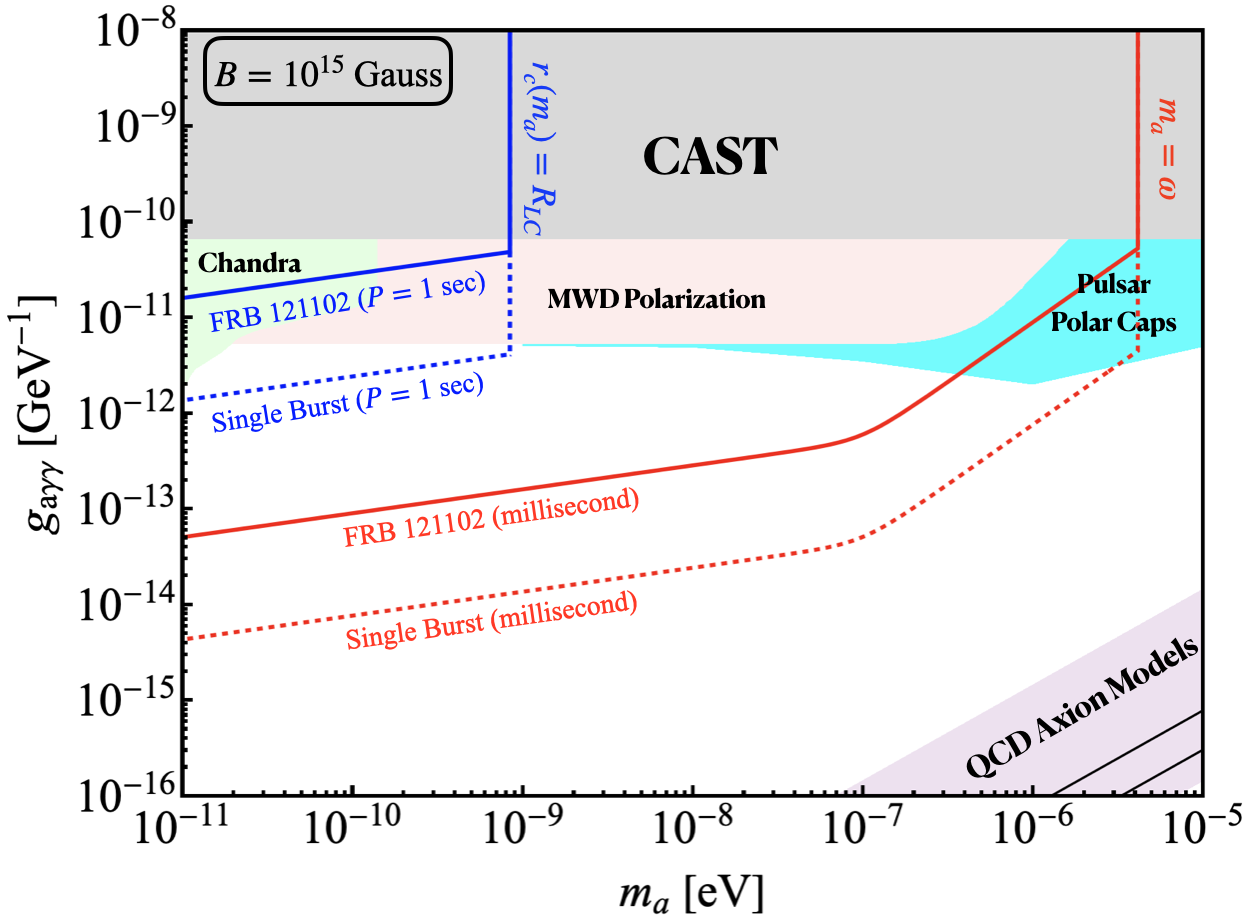}
    \caption{Minimum allowed axion-photon coupling consistent with efficiency constraints for a single burst (dashed) and from FRB 121102  (solid) assuming the source is a magnetar with surface field $B = 10^{15}$ G and rotational period $P = 1$ sec (blue) or $P = 1$ ms (red). We assume a burst of fluence $F = 1 \text{ Jy}\cdot\text{ms}$ with central frequency $\nu =$ GHz and $\Delta \nu = \nu$. For the single burst, we have assumed a distance of $D = 1$ Gpc and conservatively set $z=0$. For additional parameter values see Sec. \ref{sec:efficiency}. Parameter space excluded by other astrophysical probes is shaded (\citealt{Wouters2013,Marsh2017,Reynolds2020,Sisk2021,Dessert2022, Noordhuis2022}). QCD axion models are shaded in light purple.} 
    \label{fig:Exclusion}
\end{figure}

The radio efficiency in our model, $\eta_r = \eta_{\gamma \to a} \times \eta_{a\to \gamma}$, can be computed from (\ref{eqn:pga}) and (\ref{eqn:pag}). In Fig.~\ref{fig:Exclusion}, we show the minimum axion-photon coupling $(\gagg)$ at a given mass ($m_a$) consistent with the efficiency constraints defined above. In this plot, we assume the \alfven wave becomes charge-starved at a distance of 10 $R_\text{ns}$ and adopt \alfven wave parameters, $\omega_A = 2\pi/\text{km}$, $k_\perp = 0.1 \omega_A$, $\varepsilon = 10^{-3}$, $\gamma_0 = 10^4$. We also assume a beaming factor $f_b = 0.01$ for a single burst and $F_b = 0.1$ for the repeater. For re-conversion in the wind, we assume the pre-flare wind luminosity is enhanced by a factor of $10^2$ compared to that of the persistent wind and that the pair multiplicity factor is $\mathcal{M} = 10^2$.

\section{Discussion and Conclusions} \label{sec:conclusion}

While inner magnetar magnetosphere models of FRBs successfully explain many observed temporal and polarimetric features, general arguments suggest that FRBs are susceptible to strong scattering as they propagate through the magnetosphere (\citealt{Lyubarsky2020, Beloborodov2021}), necessitating a viable transport model. Motivated by this, we propose that FRBs may be transported out of the magnetosphere due to their mixing with axions. Ultralight axions naturally mix strongly with electromagnetism only very close to the magnetar surface and far outside the light-cylinder, where the plasma frequency in the wind coincides with the axion mass. Any inner magnetosphere production of FRBs that is accompanied by large $\epar$ necessarily produces axions in great abundance. As axions propagate through the magnetosphere, they effectively decouple from the plasma as the magnetic field drops. In the enhanced, pre-flare magnetar wind, axions encounter a plasma with monotonically decreasing plasma frequency and eventually reach a critical radius at which the axion mass coincides with the plasma mass. At this level crossing, axions mix strongly with radio photons, even in a weaker magnetic field.

We have presented a concrete model in which \alfven waves, produced by a magnetar flare, become charge starved in the inner magnetosphere and produce O modes (\citealt{Kumar2020}) and axions through a two-stream instability, though we reiterate that our mechanism is viable for any inner magnetosphere emission mechanism that hosts large $\epar$. The axions then resonantly re-convert to coherent radio waves in the wind. This model successfully explains many observational properties of FRBs. Firstly, we have shown that in this mechanism a large fraction of the flare magnetic energy can be converted to outgoing radio photons, satisfying FRB efficiency constraints. The axion burst produced in the inner magnetosphere has rapid temporal variability that is retained throughout its propagation, consistent with observations. Our scenario predicts a high degree of linear polarization, set by the magnetic field in the conversion region. Crucially, this class of models allows for FRB escape through both the open and closed magnetosphere, obviating the need for beamed emission along open field lines. 

Finally, we comment that a smoking-gun signal of axion-mediated FRBs is the presence of axion counterparts associated with FRBs. We explore the prospects for observing these counterparts with dedicated axion detection experiments in upcoming work.


\section{Acknowledgments}

I thank Anatoly Spitkovsky, Roger Blandford, Pawan Kumar, Ashley Bransgrove, Sam Witte, Sasha Philippov, Ben Safdi, and Ken van Tilburg for useful discussions. I am particularly grateful to Anatoly Spitkovsky, Sam Witte, and Ben Safdi for useful comments on earlier versions of this manuscript. This work was supported by the Princeton Center for Theoretical Science postdoctoral fellowship.

\vspace{2in}

\bibliography{sample631}{}
\bibliographystyle{aasjournal}

\end{document}